\documentclass[aps,showpacs,superscriptaddress,pra,twocolumn]{revtex4}

\def\nn{\nonumber}
\usepackage{epsfig}
\usepackage{amsmath,amssymb,amsthm}
\usepackage{graphicx}
\usepackage{psfrag}
\usepackage{bm}

\def\s2{\sigma^2}
\usepackage[dvips]{color}

\begin{document}

\title{Band inversion at critical magnetic fields in a  silicene quantum dot}

\author{E. Romera}
\affiliation{Departamento de F\'{\i}sica At\'omica, Molecular y Nuclear and
Instituto Carlos I de F{\'\i}sica Te\'orica y
Computacional, Universidad de Granada, Fuentenueva s/n, 18071 Granada,
Spain}

\author{M. Calixto}
\affiliation{Departamento de Matem\'atica Aplicada and
Instituto Carlos I de F{\'\i}sica Te\'orica y
Computacional, Universidad de Granada,
Fuentenueva s/n, 18071 Granada, Spain}

\date{\today}

\begin{abstract}

We have found out that the band inversion in a silicene quantum dot
(QD), in perpendicular magnetic $B$ and electric $\Delta_z$ fields,
drastically depends on
the strength of the magnetic field.  
We study the energy spectrum of the silicene QD where the electric field
provides a tunable band gap $\Delta$. Boundary conditions introduce chirality, so that negative and positive angular momentum $m$ 
zero Landau level (ZLL) edge states show a quite different behavior regarding the band-inversion mechanism underlying the topological 
insulator transition. We show that,  whereas some ZLLs suffer band inversion at $\Delta=0$ for any $B>0$, 
other ZLLs only suffer band inversion above  critical values  of the
magnetic field at nonzero values of the gap.

\end{abstract}

\maketitle

\section{Introduction}

It is believed that silicene opens new opportunities for electrically tunable nanoelectronic devices \cite{Padova,EzawaPRL}. The quantum
spin Hall effect \cite{liu2011}, chiral superconductivity \cite{liu2013}, giant magnetoresistance \cite{Xu2012} and other exotic 
electronic properties have been predicted in silicene. 

Silicene takes part of an emerging category of materials called ``topological insulators''. 
In these materials, the energy gap $\Delta$ between the occupied and empty states is  inverted or ``twisted'' for surface or edge states 
basically due to a strong spin-orbit interaction $\Delta_\mathrm{so}$ (namely, $\Delta_\mathrm{so}=4.2$ meV for silicene).

\begin{figure}
\includegraphics[width=6cm]{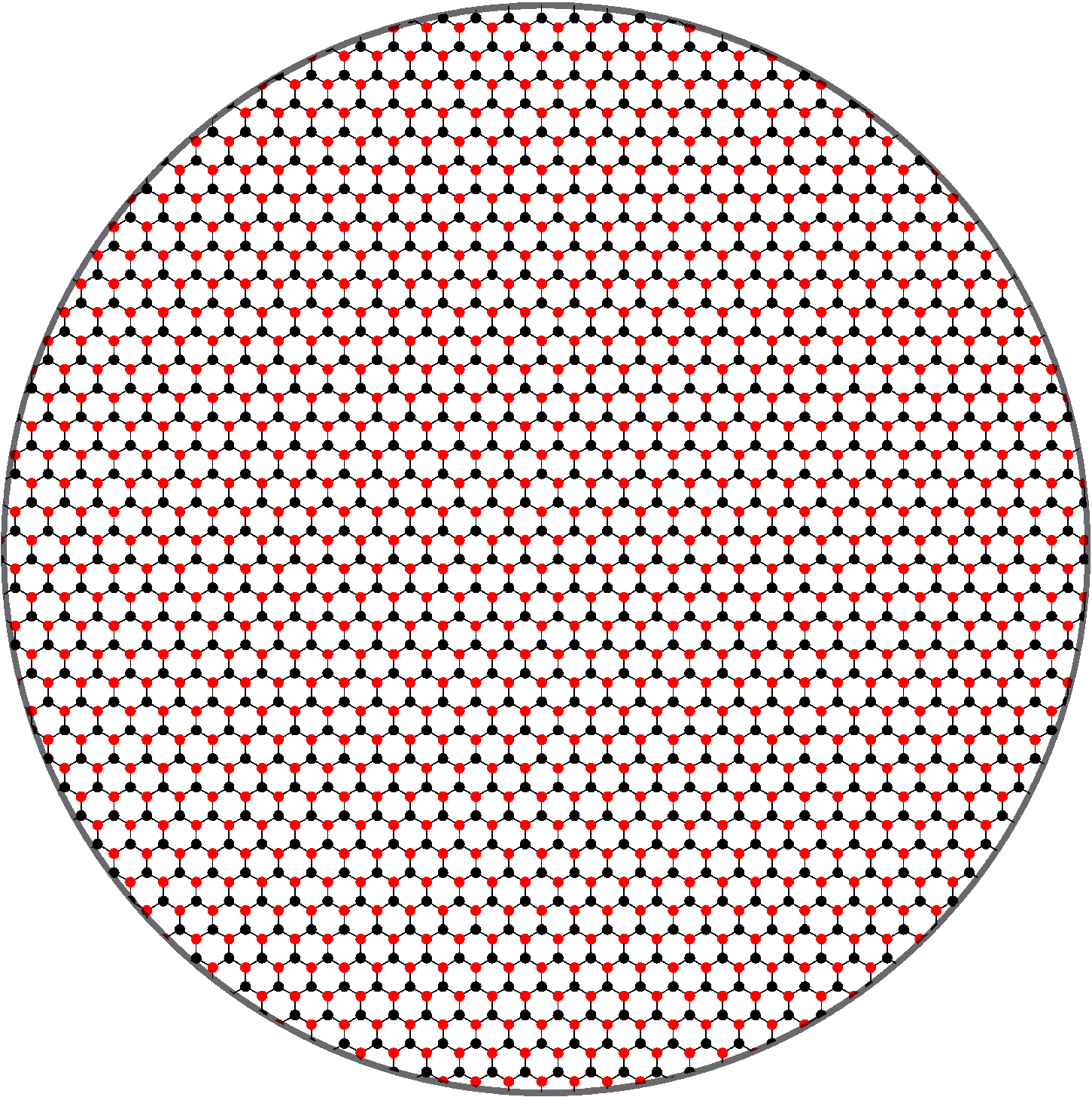}\\ 
\includegraphics[width=7cm]{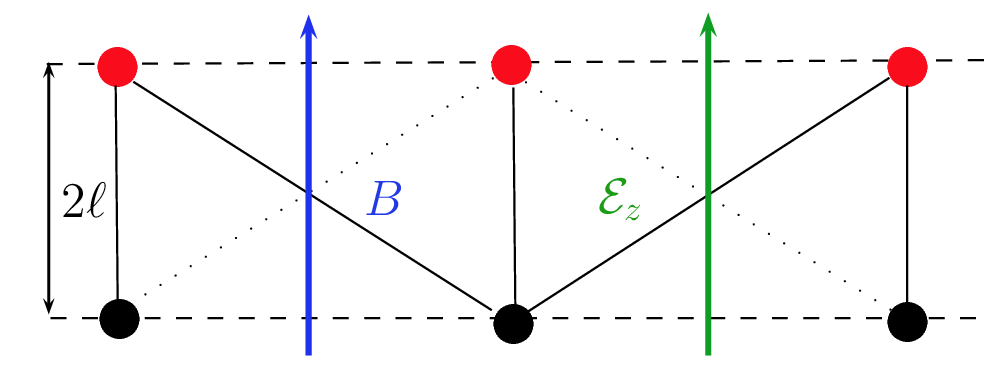}
\caption{Illustration of the buckled honeycomb lattice of a silicene quantum dot of radius $R$ in the presence of perpendicular electric $\mathcal{E}_z$ and magnetic $B$ fields.  
Red and black sites form two sublattices separated by a distance $2\ell$, with $\ell=0.22$\AA. 
The lattice distortion (compared to graphene) is due to the large ionic radius of the silicon atoms. The structure generates a staggered sublattice potential $\Delta_z=2\ell\mathcal{E}_z$ 
between red and black sites which provides a tunable band gap $\Delta$.}
\label{fig0}
\end{figure}

The low energy electronic properties of a large family of topological insulators and superconductors are well described by the Dirac equation  \cite{Shen}, in particular, 
some 2D gapped  Dirac materials isostructural with graphene like: silicene, germanene, stannene, etc. Compared to graphene, these materials display a large 
spin-orbit coupling and show quantum spin Hall effects \cite{Kane,Bernevig}. Applying a perpendicular electric field $\mathcal{E}_z$ 
to the material sheet, generates a tunable band gap (Dirac mass) $\Delta=(\Delta_z-s\xi\Delta_{\mathrm{so}})/2$, with $s=\pm 1$ the spin, $\xi=\pm 1$ the valley and 
$\Delta_z=2\ell\mathcal{E}_z$ the electric potential (see Figure \ref{fig0}). 
There is  a topological phase transition  \cite{tahir2013}  from a topological  
insulator (TI, $|\Delta_z|<\Delta_\mathrm{so}$) to a band insulator (BI, $|\Delta_z|>\Delta_\mathrm{so}$), at a charge neutrality
point (CNP)  $\Delta_z^{(0)}=s\xi\Delta_\mathrm{so}$, where there is a gap cancellation, $\Delta=0$, between the perpendicular electric field
and the spin-orbit coupling, thus exhibiting a semimetal behavior. In general, a TI-BI transition is characterized by a band inversion with
a level crossing at some critical value of a control parameter (electric field, quantum well thickness  \cite{Bernevig},
etc). { In silicene, in absence of boundary conditions, the zero Landau level (ZLL) energy is given by $E_0=-\xi\Delta$ and the band inversion at 
the CNP ($\Delta_z^{(0)}=s\xi\Delta_\mathrm{so}\Rightarrow \Delta= 0$) entails a topological phase transition. 
Actually, the transition in silicene is associated with a non-analytic
contribution to the conductivity from the zero Landau level (ZLL) topological edge or surface states, 
as a result of a band inversion. Indeed, in Ref.  \cite{tahir2013} it has been shown that the Hall conductivity jumps from $0$ to $e^2/h$ at the CNP, by tuning the electric field, 
thus reflecting the transition the from a trivial insulator to a Hall insulator at the CNP. We address the reader to Ref.  \cite{tahir2013} 
for further details on the relation between the topological phase transition in silicene and the change in the character of the ZLL at the CNP.}

Finite size and boundary conditions on the effective field theory describing these materials brings some
extra features \cite{Hasan}. In this letter we study Berry-Mondragon \cite{Berry} boundary conditions for a circular silicene QD of radius $R$, which introduces some novelties  
with regard to the previous discussion and leads to additional interesting physical 
phenomena.  In reference \cite{Schnez}  the authors studied the energy spectrum of a circular graphene QD with radius $R$ subjected to
a perpendicular magnetic field $B$. Here we consider a circular silicene QD subjected to perpendicular magnetic and electric fields, the last one providing a tunable band-gap and 
introducing new interesting physics with potential applications in nanotechnology. 

\section{Effective Hamiltonian and eigenfunctions}

The low energy dynamics of silicene in the presence of a perpendicular electric field $\mathcal{E}_z$  is described by the Dirac Hamiltonian 
in the vicinity of the Dirac points $\xi=\pm 1$ 
\begin{equation}
H_s^{\xi}=v (\sigma_x  p_x+ \xi\sigma_y  p_y )-\frac12\xi
s \Delta_{\mathrm{so}} \sigma_z+ \frac12\Delta_z \sigma_z,
\label{hamiltoniano}
\end{equation}
where ${\sigma}_j$ are the usual Pauli matrices, $s=\pm 1$ is the spin,  $v$ is the Fermi
velocity of the corresponding material (namely, $v=4.2\times 10^5$m/s for silicene),  $\Delta_{\mathrm{so}}$ and $\Delta_z$ is the spin-orbit coupling  
and $\Delta_z$ is the electric potential. We shall combine spin-orbit coupling and electric potential into the band gap $\Delta=(\Delta_z-s\xi\Delta_\mathrm{so})/2$, 
so that the explicit dependence of $H_s^{\xi}$ on the spin $s$ is masked and we can simply write $H^{\xi}$. 
We also consider a perpendicular magnetic field $B$, which is implemented through the  minimal coupling $\vec{p}\to \vec{p}+{e}\vec{A}$ for the momentum, 
where $\vec{A}=B(-y,x)/2$ is the vector potential in the symmetric gauge. Since $H^\xi$ commutes with angular momentum, in order to solve the eigenvalue problem 
$H^\xi\Psi^\xi=E\Psi^\xi$, we choose energy eigenspinors [we use polar coordinates $(r,\theta)$]
\begin{equation}
 \Psi^\xi_m(r,\theta)=e^{i\xi m\theta}[\psi_m^\xi(r), e^{i\xi\theta}\chi_m^\xi(r)]^t
\end{equation}
($t$ means transpose) which also are eigenstates of angular momentum with eigenvalue $m$ (an integer). The regular solutions at the origin are
\begin{eqnarray}
 \psi_m^\xi(r)&=&\frac{iB\pi v \hbar e^{-\frac{B \pi r^2}{4\phi}}}{(E-\Delta) 2^{\frac{m}{2}-1}r^m\phi}\times \\
 && \left[L_{-1-m\tilde{\xi}+a}^{-m}(\frac{B \pi  r^2}{\phi})-\tilde{\xi}L_{-m\tilde{\xi}+a}^{-m-1}(\frac{B \pi  r^2}{\phi})\right],\nn\\
 \chi_m^\xi(r)&=&\frac{e^{-\frac{B \pi r^2}{4\phi}}}{2^{\frac{m}{2}}r^{1+m}}L_{-m\tilde{\xi}+a}^{-1-m}(\frac{B \pi  r^2}{\phi}),
\end{eqnarray}
where $\phi=2\pi\hbar/e$ is the magnetic Dirac flux quantum, $L_n^m$ are the associated Laguerre polynomials and we are denoting by $a={(E^2-\Delta^2)\phi}/({B\pi v^2\hbar^2})$ and $\tilde{\xi}=(\xi-1)/2$. 
The Berry-Mondragon \cite{Berry} boundary condition $\chi_m^\xi(R)/\psi_m^\xi(R)=i\xi$ at radius $r=R$ provides the characteristic equation for the allowed energies $E$ of the QD.

\section{Energy spectrum: analytic and numerical study}

\begin{center}
\begin{figure}
\centering{\includegraphics[height=8cm,angle=-90]{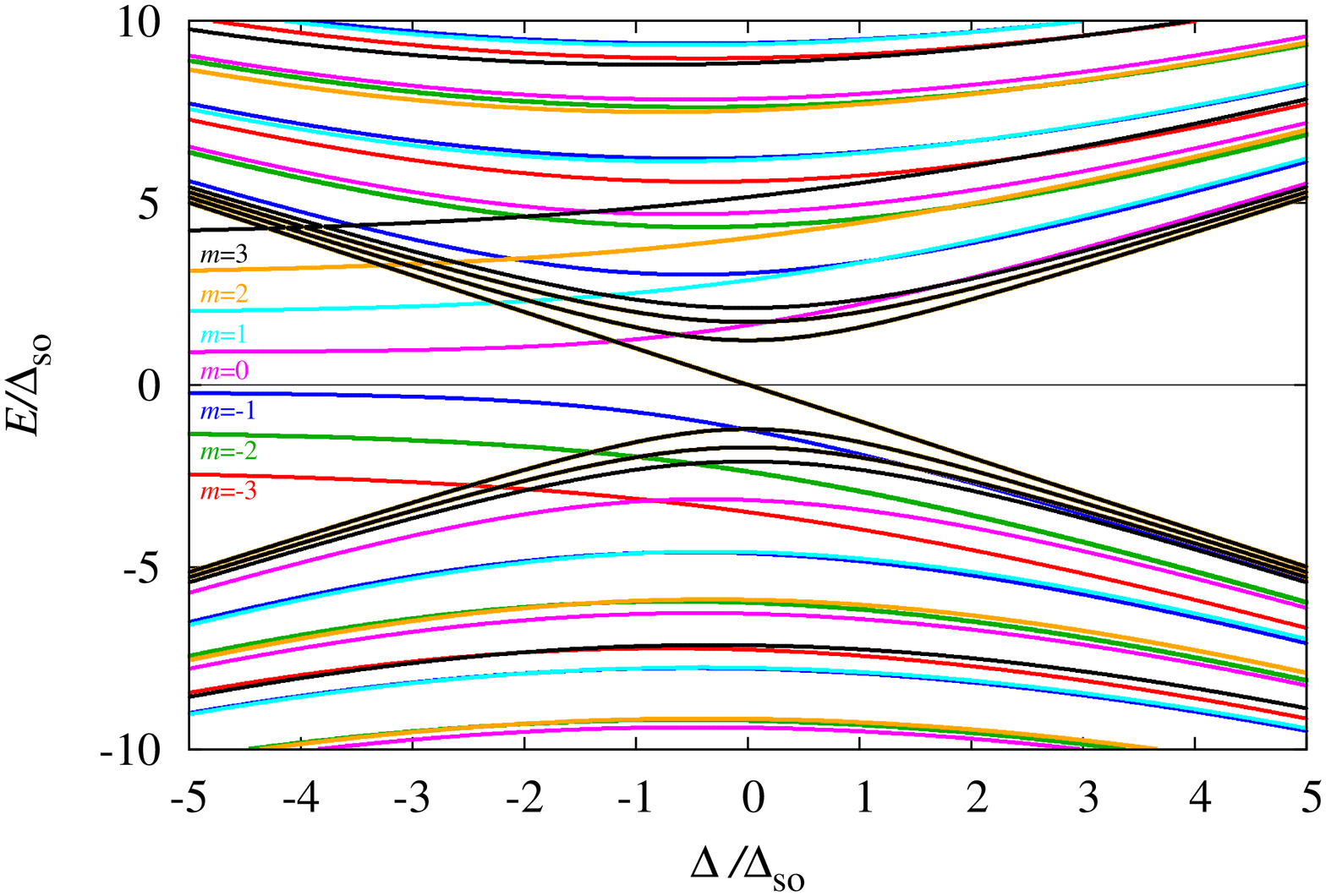}}\\
\includegraphics[height=8cm,angle=-90]{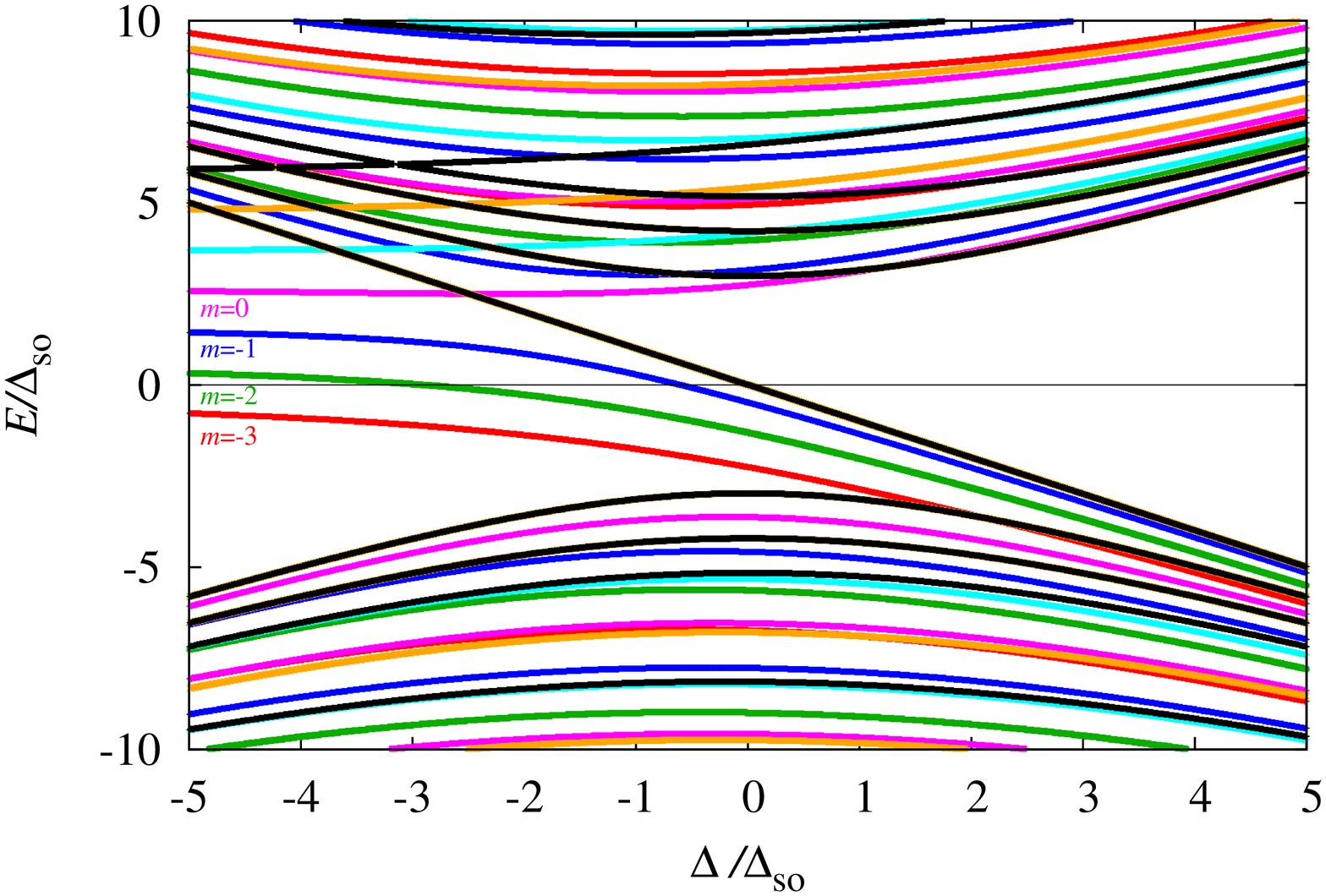}
\caption{Low energy spectrum (for angular momenta $m=-3,\dots,3$ and valley $\xi=1$) of a silicene quantum dot of radius $R=70$nm in
the presence of a  perpendicular magnetic field of $B= 0.1$T (top panel) and   $B= 0.6$T (bottom panel), bellow and above the critical value $B_c=0.136$T, respectively. 
Energy is given as a function of the gap $\Delta$, which is tunned by applying 
a perpendicular electric field. Energy and gap are measured in $\Delta_\mathrm{so}$ units.}
\label{fig2}
\end{figure}
\end{center}

We have numerically solved the Berry-Mondragon boundary condition
\begin{equation}
 \beta_m^\xi(E,\Delta,B,R)=\chi_m^\xi(R)-i\xi\psi_m^\xi(R)=0 \label{Berrycond}
\end{equation}
and computed the energy $E$ spectrum of a silicene QD of radius $R=70$nm as a function of the gap $\Delta$ for magnetic field $B=0.1$T (Figure \ref{fig2}, top panel) and 
$B=0.6$T (Figure \ref{fig2}, bottom panel). We have restricted ourselves to angular momentum $m=\pm 3, \pm 2, \pm 1, 0$ and valley $\xi=1$. 
For the valley $\xi=-1$ the results are equivalent swapping $m\to -m$ and the gap $\Delta\to -\Delta$. 

As we have commented, the topological phase transition in silicene is associated with a non-analytic
contribution to the conductivity from the zero Landau level (ZLL). In absence of boundary conditions, the  ZLL corresponds to the energy $E=-\xi\Delta$ \cite{Tsaran,Tabert2013,calixto15,romera15,jstat}, and the 
band inversion at zero gap $\Delta=0$ entails a topological phase transition.  The ZLL  still remains in the QD (note the straight diagonal line along the second and fourth quadrants 
of Figure \ref{fig2} for $\xi=1$), but boundary conditions introduce chirality, which means that positive and 
negative angular momentum $m$ states have a different behavior.  For low magnetic fields, below a critical value $B_c=\phi/(2\pi R^2)$ 
[see later on eq. \eqref{semi} for a semiclassical explanation], there only
exists a band inversion, {(that is, the  ordering of the conduction and valence bands is inverted by
the tunable band gap which depends on the  spin-orbit coupling and the electric field) }  for positive angular 
momentum ($m\geq 0$) ZLLs at $\Delta=0$; all these levels are degenerate with common energy $E=-\Delta$ at valley $\xi=1$ (see Figure \ref{fig2}). Actually, 
this can also be analytically checked by realizing that $\beta_m^1(0,0,B,R)$ vanishes only if $m\geq 0$, using properties of associated Laguerre polynomials. On the contrary, 
negative angular momentum ZLLs detach more an more from the line $E=-\Delta$ as $\Delta\to-\infty$ (large negative electric field), forming an equally spaced energy band with interlevel spacing 
of $\epsilon=\hbar v/R$  (they correspond to the energy levels labeled by negative $m$'s in Figure \ref{fig2}). 

In going from $B=0.1$T (Figure \ref{fig2}, top panel) to $B=0.6$T (Figure \ref{fig2}, botom panel) we find a band inversion of some negative angular momentum ZLLs at certain nonzero gaps $\Delta_m$ (corresponding to given negative electric fields). 
For the case $R=70$ nm, this band inversion starts for the ZLL $m=-1$ at the particular critical magnetic field $B_c=\phi/(2\pi R^2)\simeq 0.136$T. Summing-up, for a given $R$ and  for $B<B_c$, 
there only exists a band inversion for positive angular momentum ZLLs at $\Delta=0$. The situation changes  for $B>B_c$, when more and more $m<0$ ZLLs 
become conductive, $E_m>0$, at a given gap $\Delta_m<0$, for increasing values of the magnetic field. For example, as it can be appreciated in Figure \ref{fig2} bottom panel, for $B=0.6>B_c$, 
the ZLLs $m=-1$ and $m=-2$ have suffered a band inversion at certain values of $\Delta<0$ (i.e., at certain values of the electric potential $\Delta_z=2\Delta+s\xi\Delta_\mathrm{so}$). These states must contribute 
to the conductivity for these critical values of the magnetic field.
\begin{figure}
\centering{\includegraphics[width=7cm]{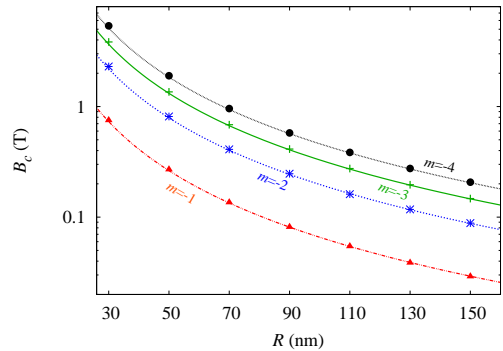}}
\caption{Critical values of the magnetic field $B_c$ (in Tesla),  as a function of the silicene QD radius $R$ (in nanometers), 
at which angular momenta ($m=$ $-1$, $-2$, $-3$ and $-4$) ZLLs suffer band inversion. 
Points correspond to the numerical results for $R=30$, $50$, $70$, $90$, $110$ and $130$ nm.  Lines correspond to the semiclassical  prediction in Eq.(\ref{semi}).}
\label{fig1}
\end{figure}

\begin{figure}
\centering{\includegraphics[height=7cm,angle=-90]{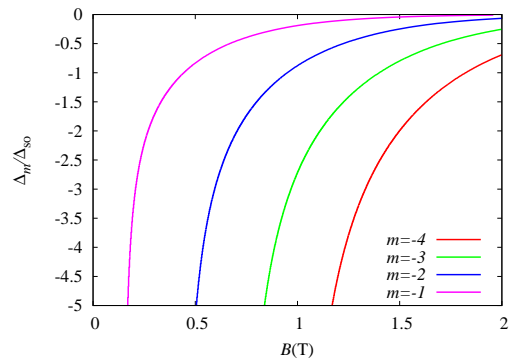}}
\caption{Gap $\Delta_m$ (in units of $\Delta_{so}$) at which there is a
  band inversion of angular momentum  ($m=-1,-2,-3$ and $-4$) ZLLs,  as a function of the  magnetic field strength $B$ (in Tesla), for a silicene QD of radius $R=70$ nm. }
\label{fig4}
\end{figure}
Let us provide a semiclassical argument that explains the aforementioned band-inversion phenomenon for negative angular momentum ZLLs and provides an analytical expression 
of the magnetic field critical values $B_c^m(R)$ at which $m<0$ ZLLs suffer a band inversion for a given QD radius $R$. Massless Dirac electrons in silicene make cyclotron motion with frequency 
$\omega=\sqrt{2}\hbar v/\ell_B$ in an external magnetic field $B$, where $\ell_B=\sqrt{\phi/(2\pi|B|)}$ is the magnetic length (the ``radius'' of the cyclotron motion for the ground state).  
The probability of finding the electron with angular momentum $m$, at a given radius $r$ in the lowest Landau level, has a sharp peak at 
$r_m=\sqrt{2|m|+1}\ell_B$. Is is clear that the corresponding circular trajectory does not fit the QD when $r_m>R$ (the QD size). This threshold provides a critical magnetic field depending on $R$ and $m$ given by
\begin{equation}
B_c^m(R)=-\xi\frac{(2m+1)\phi}{2\pi R^2},\label{semi}
\end{equation}
where we have also introduced the valley index $\xi=\pm 1$ for completeness.  Note that, as we have mentioned before, chiral symmetry is broken, which means that, for positive magnetic fields $B>0$, 
only negative (resp. positive) angular momentum $m<0$ ZLLs suffer band inversion at valley $\xi=1$ (resp. $\xi=-1$) at certain negative (res. positive) values $\Delta_m$ of the gap. 
For negative magnetic fields we have the complementary situation, according to the general formula \eqref{semi}. We have numerically checked the semiclassical formula \eqref{semi} for 
different negative angular momenta $m$ and QD radii $R$ at valley $\xi=1$. The band inversion for $m<0$ ZLLs occurs for $B>B_c^m(R)$ at certain 
negative gaps $\Delta_m(B)$ [see Figures \ref{fig2} (bottom panel) and \ref{fig4}]. At the critical point 
$B=B_c^m(R)$, we have that the energy $E_m$ of the $m<0$ ZLL vanishes only for large (negative) electric fields, that is, $\Delta_m\to -\infty$. Therefore, in order to check the prediction \eqref{semi}, we have 
numerically solved the boundary condition \eqref{Berrycond} for large (negative) electric potentials 
and several values of $m$ and $R$. The numerical results (points) in Figure \ref{fig1} confirm the semiclassical prediction (lines) in Eq. \eqref{semi}  with high accuracy for several $m$ and $R$.

For a given QD size (for example $R=70$nm) we have found out that, when the value of the magnetic field increases, 
there are more and more band inversions of $m<0$ ZLLs at certain gaps $\Delta_m<0$ (see Figure \ref{fig4}), corresponding to increasing  values of $|m|$ with $m<0$. Moreover, for a given $m<0$,
$\Delta_m$ goes to zero as $B$ increases. This calculation has been done by numerically solving the boundary condition \eqref{Berrycond} for $E=0$ and valley $\xi=1$. 
We have illustrated  this result in Figure \ref{fig4}  for angular momentum ZLLs $m=-1$, $-2$, $-3$, $-4$ and for a silicene QD of radius $R=70$ nm.

\section{Conclusions}

We have studied the energy spectrum of a silicene QD of radius $R$ in the presence of perpendicular magnetic $B$ and electric $\Delta_z$ fields, the last one providing a tunable band gap $\Delta$. 
We have established the existence of critical magnetic fields, given by the semiclassical formula $B_c^m(R)=-\xi{(2m+1)\phi}/({2\pi R^2})$, above which 
angular momentum $m$ ZLLs  of a silicene QD of radius $R$ suffer band inversion and contribute to the conductivity. Boundary conditions introduce chirality, thus distinguishing positive and negative angular momentum 
edge states. When sign$(m)=$sign$(B\xi)$, all angular momentum 
ZLLs are degenerate, with energy $E=-\xi\Delta$,  and all of them suffer a band inversion at gap $\Delta=0$ for any value of the magnetic field. When sign$(m)=-$sign$(B\xi)$ [matching 
the formula for $B_c^m(R)$], the degeneracy is broken and a band inversion occurs at nonzero gap $\Delta_m(B)$ for $|B|>|B_c^m(R)|$. As $B$ increases, more and more angular momenta $m$ ZLLs suffer band 
inversion at gaps $\Delta_m(B)$, which go to zero as $B$ increases.

{We have performed our calculations in the continuous model, which is a long-wave approximation of the more fundamental tight binding model. 
Therefore, we have disregarded the effect of lattice termination on the energy spectrum \cite{PRB91}. Nevertheless, 
we hope that the effect of the boundary irregularities on the spectrum is negligible when the radius $R$ is much larger than the lattice constant, 
and our results on band inversion at critical magnetic fields remain valid at least in this regime. Of course, a more detailed calculation inside the tight binding framework  
with more realistic edge termination, like the one done in Ref. \cite{PRB91} for silicene in magnetic field, 
is necessary to account for the robustness of the band inversion phenomenon. We think that this question deserves a separate study and will be considered elsewhere.}

Anyhow, we believe that these critical phenomena in a silicene QD can lead to interesting nanotechnological applications.

\section*{Acknowledgments}

  The work was supported by 
the  Spanish projects:  MINECO FIS2014-59386-P and the Junta de Andaluc\'{\i}a projects FQM.1861 and FQM-381. {We thank the anonymous referee 
for bringing to our attention Ref. \cite{PRB91}.}


\begin{thebibliography}{99}
\bibitem{Padova} P. De Padova et al., 
J. Phys. Condens. Matter {\bf 24}, 223001 (2012).
\bibitem{EzawaPRL} M. Ezawa, Phys. Rev. Lett. {\bf 109}, 055502 (2012); 
 M. Ezawa, Monolayer Topological Insulators: Silicene, Germanene and Stanene, arXiv:1503.08914
\bibitem{liu2011} C. C. Liu, W. Feng, and Y. Yao, Phys. Rev. Lett. {\bf 107}, 076802 (2011).
\bibitem{liu2013} F. Liu,C-C Liu, K. Wu, F. Yang and Y. Yao,  Phys. Rev. Lett. {\bf 111}, 066804 (2013).
\bibitem{Xu2012} C. Xu, et al.,  Nanoscale
{\bf 4}, 3111-3117 (2012).
\bibitem{Shen} Shun-Qing Shen, {\it Topological Insulators: 
Dirac Equation in Condensed Matters},  Springer-Verlag Berlin Heidelberg 2012.
\bibitem{Kane} Kane C L and Mele E J,  Phys. Rev. Lett. {\bf 95}, 226801 (2005)
\bibitem{Bernevig} B. Andrei Bernevig, Taylor L. Hughes and Shou-Cheng Zhang, Science {\bf 314}, 1757-1761 (2006).  
\bibitem{tahir2013} M. Tahir, U. Schwingenschl\"ogl, Scientific Reports, {\bf
  3}, 1075 (2013).
  \bibitem{Hasan} M. Z. Hasan and C. L. Kane, Rev. Mod. Phys. {\bf 82}, 3045-3067 (2010)
\bibitem{Berry}  M. V. Berry and R. J. Mondragon, Proc. R. Soc. London, Ser. A
{\bf 412}, 53 (1987).
\bibitem{Schnez} S. Schnez, K. Ensslin, M. Sigrist and T. Ihn, Phys. Rev. B {\bf 78}, 195427 (2008).
\bibitem{Tsaran} V. Yu. Tsaran and S. G. Sharapov, Phys. Rev. B {\bf 90}, 205417 (2014)
\bibitem{Tabert2013} C.J. Tabert and E.J. Nicol, Phys. Rev. Lett. {\bf 110}, 197402 (2013); C.J. Tabert and E.J. Nicol, Phys. Rev. B {\bf 88},
085434 (2013).
\bibitem{calixto15} M. Calixto and E. Romera, EPL {\bf 109},  40003 (2015)
\bibitem{romera15} E. Romera and M. Calixto, J. Phys.: Condens. Matter {\bf 27}, 175003 (2015) 
\bibitem{jstat} M. Calixto and E. Romera, J. Stat. Mech. (2015) P06029. doi:10.1088/1742-5468/2015/06/P06029
\bibitem{PRB91} {P. Rakyta, M. Vigh, A. Csord\'as and J. Cserti, Phys. Rev. B {\bf 91}, 125412 (2015). doi: 10.1103/PhysRevB.91.125412}


\end{thebibliography}
\end{document}